# Further exploration of binding energy residuals using machine learning and the development of a composite ensemble model

I. Bentley [*] and J. Tedder
*Department of Physics, Florida Polytechnic University, Lakeland, Florida 33805, USA*

M. Gebran
*Department of Chemistry and Physics, Saint Mary's College, Notre Dame, Indiana 46556, USA*

A. Paul
*The Institute for Experiential AI, Northeastern University, Boston, Massachusetts 02115, USA and CDNM, Brigham and Women's Hospital, Harvard Medical School, Boston, Massachusetts 02115, USA*

This paper describes the development of the Four Model Tree Ensemble (FMTE). The FMTE is a composite of machine learning models trained on experimental binding energies from the Atomic Mass Evaluation (AME) 2012. The FMTE predicts binding energy values for all nuclei with $N > 7$ and $Z > 7$ from AME 2020 with a standard deviation of 76 keV and a mean average deviation of 34 keV. The FMTE model was developed by combining three new models with one prior model. The new models presented here have been trained on binding energy residuals from mass models using four machine learning approaches. The models presented in this work leverage shape parameters along with other physical features. We have determined the preferred machine learning approach for binding energy residuals is the least-squares boosted ensemble of trees. This approach appears to have a superior ability to both interpolate and extrapolate binding energy residuals. A comparison with the masses of isotopes that were not measured previously and a discussion of extrapolations approaching the neutron drip line have been included.

## I. INTRODUCTION

Traditional microscopic and micro-macro nuclear binding energy models typically reproduce experimental values with standard deviations ranging from about 200 to 700 keV [1]. More accurate predictions of nuclear binding energies are important for the planning of new experimental measurements (e.g., at Argonne National Laboratory [2], Michigan State University [3], and the University of Jyväskylä [4]). Additionally, binding energy values are critical inputs into astrophysical calculations (see, e.g., *r*-process sensitivity studies [5–7]). It has been suggested that a maximum standard deviation of 50 keV is needed for these astrophysical calculations [8]. Additionally, the choice of model has an impact on *r*-process results because mass models diverge far from stability [9].

Machine Learning (ML) provides an alternative approach to determine binding energies. ML approaches have been used to directly model binding energy using a Neural Network (NN) [10], Support Vector Machine (SVM), and Gaussian Process Regression (GPR) [11], all of which achieve a comparable level of accuracy in reproducing experimental values. Recently, mass models have been leveraged as a starting point, with ML providing corrections to residuals based on deep NNs [12,13] and convolutional NNs [14], and using tree-based ML [15]. Generally, these approaches have been successful in reproducing experimental binding energy values achieving standard deviations below 200 keV.

In this paper, we continue using the general methodology from our prior work [16]. The methodology involves training and testing four ML approaches to model binding energy residuals from a selected set of mass models. The ML approaches used are SVM, GPR, fully Connected Neural Network (FCNN), and the Least-Squares Boosted Ensemble of Trees (LSBET). In an attempt to minimize overfitting, we train the data using cross-validation of a subset of the Atomic Mass Evaluation (AME) 2012 [17] data and determine the preferred model from each approach based on an independent test set of different isotopes using AME 2020 [18] data.

This paper describes the development of 16 models resulting from combining ML with binding residuals from the selected mass models. In the interest of determining a preferred model for future use, we have compared these models with the 12 models from Ref. [16]. This paper also includes a discussion of how a composite model comprised of several of these models was constructed and how well it performs on the basis of new experimental measurements.

Section II introduces the mass models that have been selected and discusses how the residuals are determined. This section also discusses the discrepancies between the original mass models, which motivates the development of a new approach. Section III describes the generation of testing and training sets. Section IV describes the ML approaches used to generate the new models. Section V describes the physical features that are used to train the models. Section VI A discusses the intuition that can be gained by calculating the global Shapley values for each model. Section VI B discusses the development of a composite model made from four LSBET models. Sections VI C, VI D, and VI E contain the analyses using Garvey-Kelson relationships, comparisons of the models based on recent mass measurements, and a test of the extrapolation of the models for neutron-rich nuclei, respectively.

[*]Contact author: ibentley@floridapoly.edu

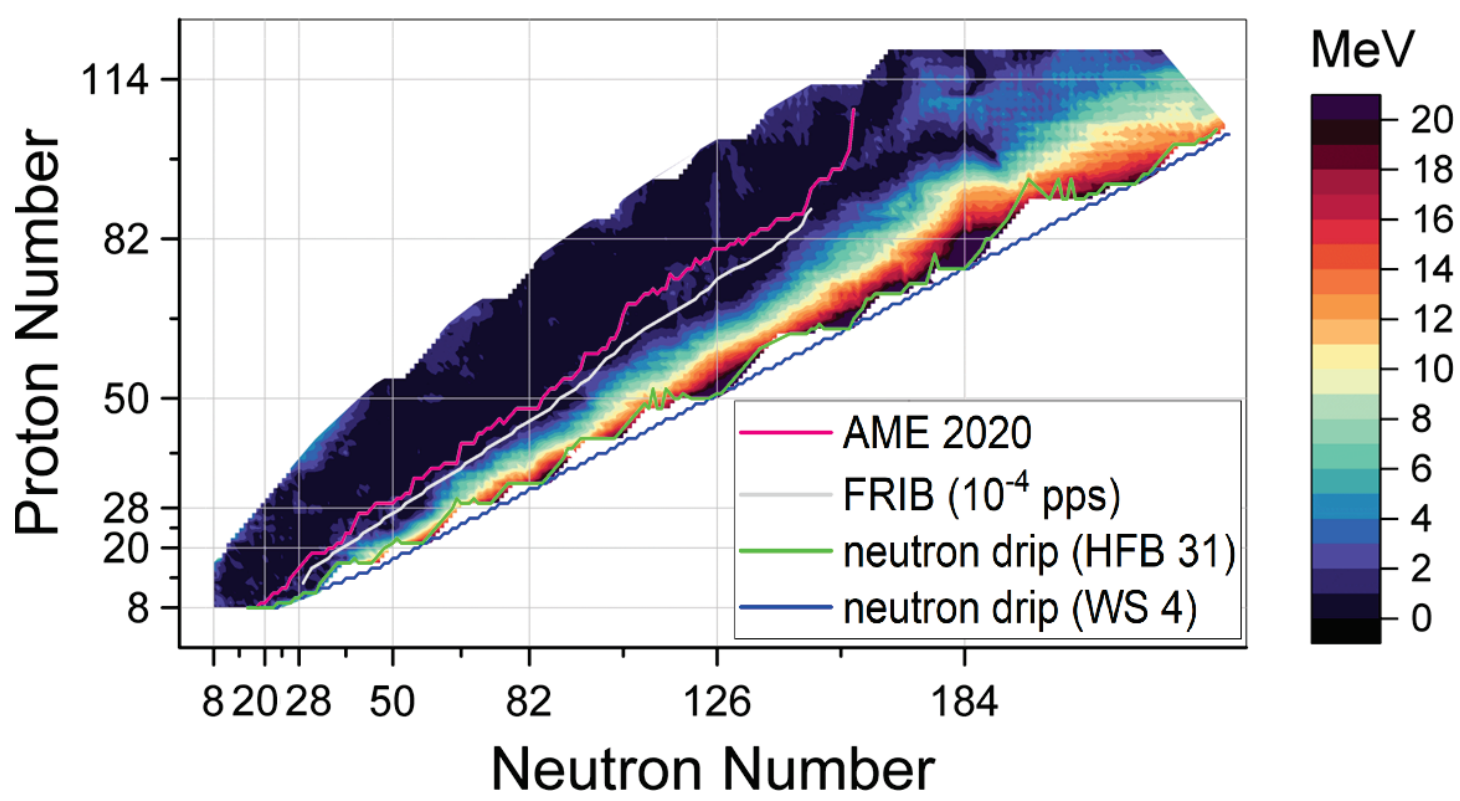

FIG. 1. Absolute difference between binding energies from HFB 31 [21] and WS 4 [22] in MeV. Lines have been included to denote the neutron-rich AME 2020 boundary, the FRIB $10^{-4}$ pps line [23], and drip lines determined using binding energies from both mass models.

## II. MASS MODELS

Mass models can be macroscopically motivated, often based on a charged liquid droplet of nuclear matter. Mass models can also be microscopic, utilizing nucleonic potentials and accounting for nucleon-nucleon interactions. Lastly, they can be a combination of the above with mixed considerations. In our previous work [16], we used two simple liquid-droplet-based models and the 28-parameter Duflo-Zuker (DZ) model [19].

For this analysis, we have selected models that were all published after the AME 2012 was released. They are the Finite-Range Droplet Model (FRDM) 2012, which is based on a liquid droplet with microscopic corrections [20], Hartree-Fock-Bogoliubov (HFB) 31, which is based on microscopic interactions and pairing along with phenomenological terms [21], and Weizsäcker-Skyrme (WS) 4, which is a micro-macro model inspired by Skyrme energy density functionals and a liquid-drop model [22].

Mass models like FRDM, HFB, and WS generally differ by less than 1 MeV in regions where binding energies have been experimentally measured. Figure 1 demonstrates the difference between HFB and WS models.

New measurements at the Facility for Rare Isotope Beams (FRIB) at Michigan State University can be used as an estimate of what is on the horizon for new measurements. We have used FRIB's production calculator [23] to determine the $10^{-4}$ pps boundary, as used in Ref. [24] as a reasonable production limit estimate for FRIB. Between the neutron-rich boundary of AME 2020 and this FRIB limit, the difference between these two models remains generally small, near 1 MeV.

As the neutron excess increases, the deviations between these two models grow, reaching more than 20 MeV. One consequence of this deviation is that the neutron drip lines as measured for a given neutron number are separated often by nearly 2.5 protons on average.

In our prior work, we found that the machine learning models often resulted in binding energy values that were more similar to each other than the values in the original models [16]. One aspect of interest is to investigate whether this convergence among ML models persists.

We have also included the WS 4 with Radial Basis Function (WSRBF) correction [22,25]. The WSRBF has been included for benchmarking purposes to see what can be gained from training models on an already-modeled residual.

The binding energy residuals for each model where determined by removing the model value from experiment, such that

$$\Delta B_{\text{model}}(N, Z) = B_{\text{expt}}(N, Z) - B_{\text{model}}(N, Z). \qquad (1)$$

The $\Delta B$ values have been included in Fig. 2. The FRDM has on average the largest binding energy residuals, which are occasionally more than 2 MeV. Meanwhile, the WSRBF values are often less than 250 keV in magnitude as demonstrated in Fig. 2(d).

## III. TESTING AND TRAINING DATA

A critical aspect of our methodology is the use of robust independent datasets for training and testing. We have used the same training and test sets from Ref. [16].

The training set is the AME 2012 [17] with 400 isotopes removed. The isotopes removed include 57 values that have changed by more than 100 keV between AME 2012 and AME 2020 [18]. Another 17 values were removed because they were marked as measured in AME 2012 but replaced by extrapolated values in AME 2020. Lastly, the final 326 values were excluded so that the test set could be sufficiently large (approximately 25% of the training). These were selected to be one out of every seven of the remaining isotopes. These seeded values mainly allow for interpolation to be tested, while the remaining values, which often lie at the extremes, serve as a measure of nearby extrapolation, as can be seen in Fig. 3.

There were 121 measurements for new isotopes in AME 2020 that were not included in AME 2012 shown in Fig. 3. These 121 measurements, along with the 57 substantially changed values and the 326 seed values, were combined to create 504 binding energies used in our test set. This test set will be used to identify the best model, based on which model has the lowest mean average deviation ($\overline{AE}$) for each pair of mass model and ML approach.

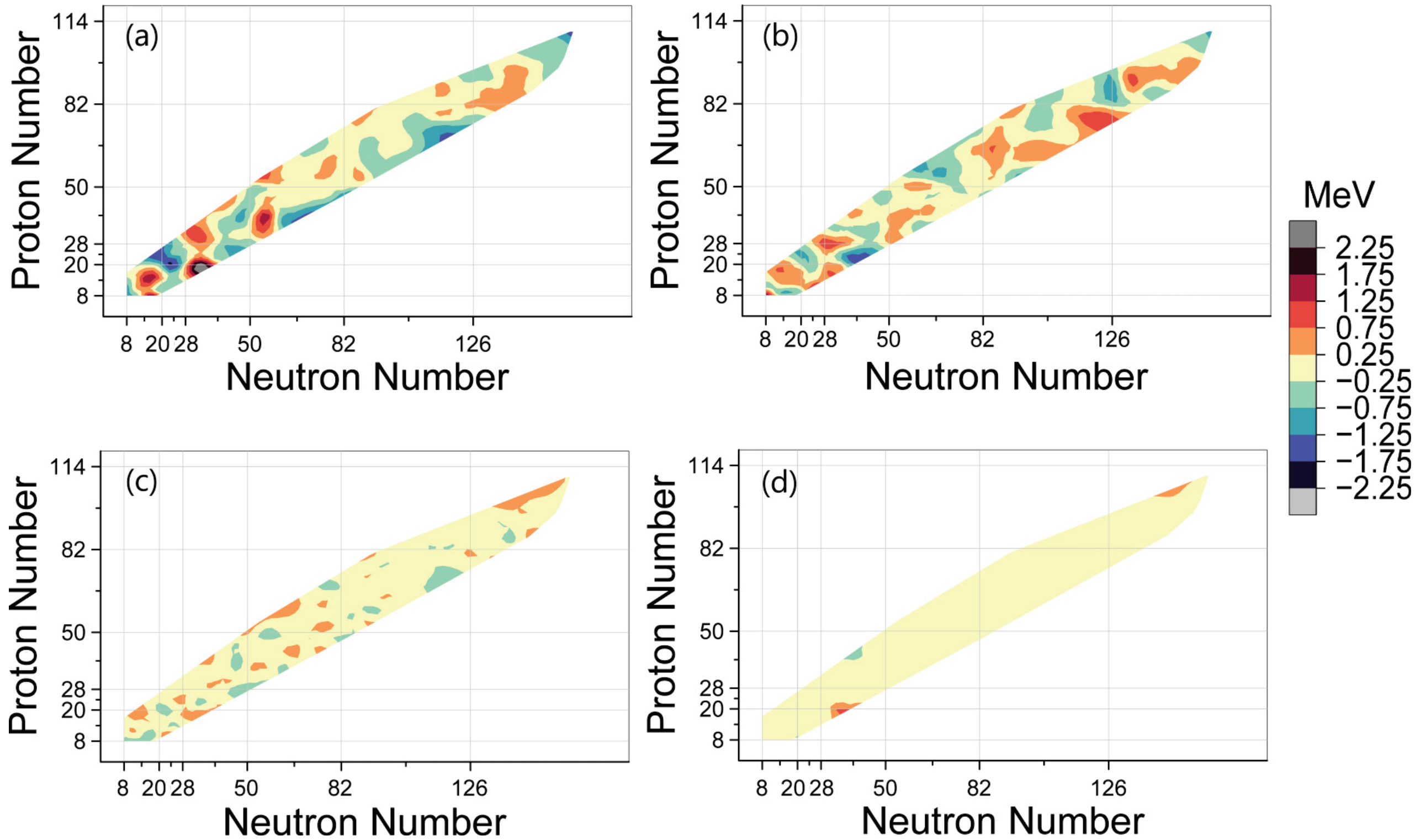

FIG. 2. The binding energy residuals (a) $\Delta B_{\mathrm{FRDM}}$, (b) $\Delta B_{\mathrm{HFB}}$, (c) $\Delta B_{\mathrm{WS}}$, and (d) $\Delta B_{\mathrm{WSRBF}}$ for the four theoretical models compared with experimental measurements from AME 2020.

The mean experimental uncertainties are 20 keV for the training set, 26 keV for the full AME 2012, 44 keV for the test set, and 23 keV for the full AME 2020 [16]. These values have been provided to give context for the model evaluation discussion.

## IV. MACHINE LEARNING

ML models were generated for each of the mass models described in Sec. II. The following general categories of machine learning approaches were used to model the residuals resulting from Eq. (1): SVM [26,27], GPR [28], FCNN [29,30], and LSBET [31,32].

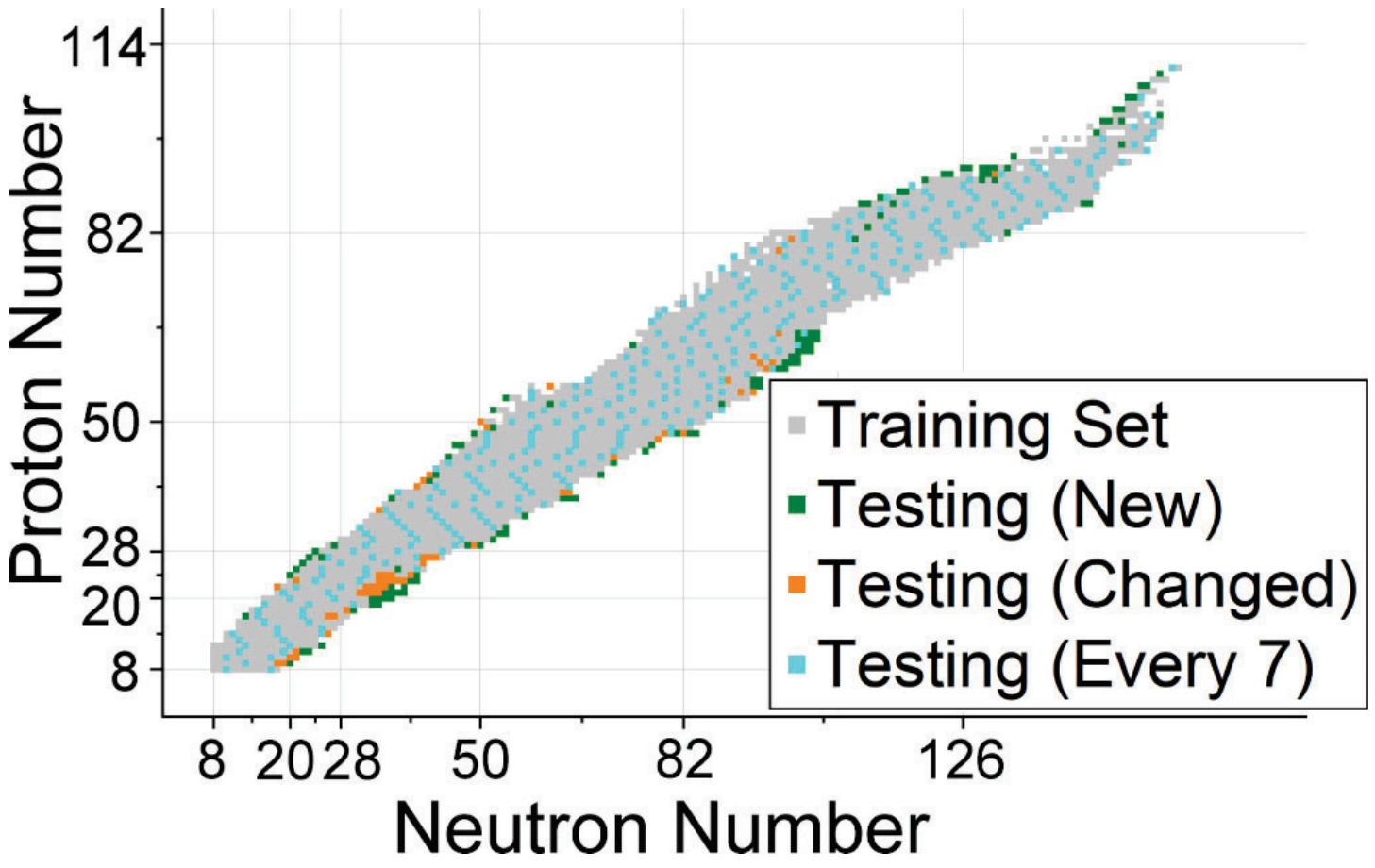

FIG. 3. The training set comprised of AME 2012 data along with the three constitutes of the testing set from AME 2020, specifically new values, changed values, and the selection for one out of every seven throughout. Reproduced from Ref. [16].

All the machine learning training employed a fivefold cross-validation scheme to mitigate the risk of overfitting. We used Bayesian optimization with the mean squared error as the primary loss function for between 50 and 250 iterations. The acquisition function used for hyperparameter tuning was *expected improvement per second plus*.

### A. Support vector machine

SVM regression [26] is a machine learning approach that we have used to model binding energy residuals. SVM regression involves the use of a kernel to determine patterns in the data. Linear, quadratic, cubic, and Gaussian kernels were all tested.

Similar to prior investigations (see e.g., Refs. [11,16]), we have found that the Gaussian kernel of the form

$$K_G(x_i, x_j) = e^{-\gamma r^2} \tag{2}$$

generates the best results. Here $\gamma$ is the kernel coefficient, which determines how much influence each training example has, with higher values creating more localized influence and lower values promoting smoother generalization. The variables $x_i$ and $x_j$ represent two data points, and $r$ is the Euclidean distance between the two points:

$$r = ||x_i - x_j|| = \sqrt{(x_i - x_j)^T (x_i - x_j)}. \tag{3}$$

The SVM regression optimization problem incorporates three key hyperparameters that must be tuned. The first is the kernel scale parameter $\gamma$ that appears directly in the Gaussian kernel equation above. The remaining two hyperparameters, the box constraint $C$ and epsilon parameter $\epsilon$, appear in the SVM formulation itself rather than the kernel function.

The box constraint $C$ controls the penalty imposed on observations with large residuals. Higher $C$ values enforce stricter adherence to training data, but may lead to overfitting, while lower values promote generalization. The $\epsilon$ parameter governs the margin of tolerance where no deviations greater than $\epsilon$ contribute to the loss function, with larger $\epsilon$ values creating wider tolerance margins and potentially fewer support vectors. Together with the kernel scale $\gamma$, these three hyperparameters determine the model's complexity and generalization behavior.

### B. Gaussian process regression

Gaussian Process Regression (GPR) [28] is a nonparametric machine learning approach that offers several advantages for our modeling task. GPR naturally adapts to the complexity of the underlying data without requiring a fixed number of model parameters, and it provides uncertainty quantification alongside predictions. Our GPR implementation utilizes kernel functions to measure similarity between the training and prediction points specified in Eq. (3). This enables trained GPR models to capture diverse patterns in the data, from smooth trends to complex nonlinear structures.

We evaluated GPR models using four possible kernel functions, each with different characteristics regarding smoothness and flexibility. The tested kernels include the exponential kernel, the squared exponential kernel, the rational quadratic kernel, and the Matérn 5/2 kernel. Each kernel encodes different assumptions about the underlying function and handles complex data patterns differently.

The exponential kernel has the form

$$K_{\mathrm{exp}}(x_j, x_k) = \sigma_f^2 e^{\left(-\frac{r}{\sigma_l}\right)}, \tag{4}$$

where $\sigma_l$ is the characteristic length scale that controls how quickly the correlation decays with distance, and $\sigma_f^2$ is the signal variance that determines the overall scale of function variations. This kernel produces functions that are continuous but not differentiable, making it suitable for modeling rough or nonsmooth data. In our analysis, this kernel was used exclusively by the WSRBFGPR model, where its simplicity was sufficient because complex structures were already captured by the radial basis function correction.

There were two kernels tested, but they were not used by any best GPR-based model, one of which includes the squared exponential kernel. This kernel is given by

$$K_{\mathrm{se}}(x_j, x_k) = \sigma_f^2 e^{\left(-\frac{r^2}{2\sigma_l^2}\right)}. \tag{5}$$

The squared exponential kernel generates infinitely differentiable functions and is appropriate for modeling very smooth data with strong local correlations.

The other kernel not used by any best GPR-based model, the rational quadratic kernel, is given by the form

$$K_{\mathrm{rq}}(x_j, x_k) = \sigma_f^2 \left(1 + \frac{r^2}{2\alpha\sigma_l^2}\right)^{-\alpha}, \tag{6}$$

where $\alpha$ is a positive parameter that controls the relative weighting of large-scale and small-scale variations. This kernel can be viewed as a mixture of squared exponential kernels with different length scales.

Three of our best-performing GPR models adopted the Matérn 5/2 kernel. This kernel provides a good balance between flexibility and smoothness. It is defined as

$$K_{\mathrm{M5/2}}(x_j, x_k) = \sigma_f^2 \left(1 + \frac{\sqrt{5}r}{\sigma_l} + \frac{5r^2}{3\sigma_l^2}\right) e^{\left(-\frac{\sqrt{5}r}{\sigma_l}\right)}. \tag{7}$$

The Matérn 5/2 kernel generates functions that are twice differentiable, making it more flexible than the exponential kernel and less restrictive than the squared exponential kernel. This characteristic makes it particularly suitable for modeling physical phenomena that exhibit moderate smoothness.

There are four hyperparameters optimized during the training of the best GPR models. The first is the basis function (zero, constant, or linear), which determines the background behavior of the function. Next is the characteristic length scale $\sigma_l$, which controls the rate of correlation decay. Third is the signal variance $\sigma_f^2$, which scales the overall function variation. The final kernel-specific parameter, $\alpha$, used in the rational quadratic kernel, was also optimized when applicable.

### C. Fully connected neural network

We also trained FCNN models on the binding energy residuals. Artificial neural networks are computational models inspired by the structure of biological neural networks, where nodes (analogous to neurons) are connected by weighted links (analogous to synapses) [29]. In FCNNs, each node in a given layer is connected to every node in the subsequent layer, creating a dense network architecture that can learn complex nonlinear mappings through the combination of weights, biases, and activation functions [30].

Our FCNN implementation employed several key design choices and hyperparameters. In each, we used a batch size of 128 for training, which provides a balance between computational efficiency and gradient estimation stability. Three different activation functions were systematically tested: the Rectified Linear Unit (ReLU), sigmoid, and hyperbolic tangent (tanh) functions are defined as

$$\mathrm{ReLU}(x) = \max(0, x), \tag{8}$$

$$\mathrm{sigmoid}(x) = \frac{1}{1 + e^{-x}}, \text{ and} \tag{9}$$

$$\tanh(x) = \frac{e^x - e^{-x}}{e^x + e^{-x}}, \tag{10}$$

where $x$ is the combination of inputs with their determined weights and bias. Each activation function introduces different nonlinear characteristics to the network.

Among these activation functions, tanh consistently yielded the best performance across all FCNN models, consistent with the findings reported in Ref. [16]. The tanh function's symmetric output range $[-1, 1]$ and zero-centered nature likely contribute to more stable gradient flow during backpropagation compared to sigmoid, while avoiding the potential gradient sparsity issues associated with ReLU in this application.

We systematically explored different network architectures by varying both the number of hidden layers and the number of nodes per layer. The tested configurations included networks with one, two, and three hidden layers. For each layer count, we examined various layer sizes. In two-layer networks, [10, 10], [100, 100], [200, 200], [300, 300], and [400, 400] were used, and in three-layer networks, [10, 10, 10], [100, 100, 100], [200, 200, 200], [300, 300, 300], and [400, 400, 400] were used.

Architectural optimization revealed that two-hidden-layer networks consistently outperformed single-layer and three-layer alternatives across all mass models. This suggests that two layers provide sufficient representational capacity for capturing the underlying physics while avoiding the overfitting and training difficulties often associated with deeper networks. The optimal layer size varied depending to the specific model. For the HFBFCNN model, the configuration with 400 nodes in each hidden layer achieved the best performance, while for all other models, 200 nodes per hidden layer proved optimal.

To prevent overfitting and improve generalization, we implemented $L_2$ regularization. This technique adds a penalty term to the loss function:

$$L_{\text{total}} = L_{\text{original}} + \lambda \sum_i w_i^2, \tag{11}$$

where $L_{\text{original}}$ is the primary loss function, $\lambda$ is the regularization strength hyperparameter, and $w_i$ represents the network weights. The $L_2$ penalty encourages smaller weight values, reducing model complexity and improving the network's ability to generalize to unseen data. The regularization parameter $\lambda$ was optimized during the hyperparameter tuning process for each model configuration.

### D. Least squares boosted ensemble of trees

The LSBET approach uses multiple decision trees each trained on different subsets of data, allowing the algorithm to follow an iterative procedure. At each step, a new decision tree is fitted to the residuals from the current ensemble, updating the existing ensemble by adding the predictions from the new decision tree. The new predictions are scaled to the existing ensemble by a learning rate parameter $\eta$. Lastly, the mean squared error between the predicted and true values is minimized.

The LSBET models were trained with varying numbers of learners, specifically 1000, 2000, 3000, 4000, and 5000 learners. The models trained with 3000 learners performed best in two of the four cases. In the other two cases, the relative improvement achieved with 5000 learners was negligible. Considering these findings, all the models presented using LSBET will incorporate 3000 learners.

The optimized hyperparameters for the LSBET models are the minimum leaf size and the learning rate. The number of leaves indicates the number of data observations that a leaf node must have.

## V. PHYSICAL FEATURES

Each machine learning approach utilizes physical features to train with. Ten of the physical features used to train models have to do with the number of protons and/or neutrons. Four of the physical features are directly related to the number of particles in each isotope. These are the proton number ($Z$), neutron number ($N$), mass number ($A = N + Z$), and isospin projection ($T_Z = (N - Z)/2$. The next two parameters are the shell scaling parameters from Ref. [33]:

$$\nu = \frac{2N - N_{\text{max}} - N_{\text{min}}}{N_{\text{max}} - N_{\text{min}}} \tag{12}$$

and

$$\zeta = \frac{2Z - Z_{\text{max}} - Z_{\text{min}}}{Z_{\text{max}} - Z_{\text{min}}}, \tag{13}$$

where the minimum value of $-1$ at the beginning of a shell and maximum value of 1 at a closed shell are defined by the nearest magic numbers, and a value of zero occurs if the nucleus is in the middle of a shell. These are based on the following magic numbers,

$$N_{\text{min / max}} = [2, 8, 20, 28, 50, 82, 126, 196] \tag{14}$$

and

$$Z_{\text{min / max}} = [2, 8, 20, 28, 50, 82, 114, 124], \tag{15}$$

that result from Nilsson levels from Ref. [34].

The same Nilsson levels from Ref. [34] are used to define the neutron and proton subshell numbers, labeled $N_S$ and $Z_S$, which determine which subshell level is occupied by the valence neutrons and protons, starting with 1 for the $1s_{1/2}$ orbital, 2 for the $1p_{3/2}$ orbital, 3 for the $1p_{1/2}$ orbital, and so on.

The features $N_E$ and $Z_E$ are Boolean operators indicating if the neutron or proton number is even (resulting in a value of 1) or odd (resulting in 0).

Each of the mass models used in this analysis has three or four additional shape parameters that will also be used as physical features in machine learning training. For the FRDM, the deformation parameters $\beta_2$, $\beta_3$, $\beta_4$, and $\beta_6$ are determined for the model. For HFB, it is the deformation parameters $\beta_2$ and $\beta_4$, and the charge radius ($R_C$), which accompany the binding energy values, and for the WS-based models, the deformation parameters $\beta_2$, $\beta_4$, and $\beta_6$ are provided.

It should be noted that there is some potential redundancy when using $\nu$ and $\zeta$ with $\beta_2$. The value of $\beta_2$ is known to generally have a low value near a closed shell, and it increases to a high value when near midshell in both protons and neutrons while the parameters $\nu$ and $\zeta$ measure the location relative to shell closures. Figure 4 demonstrates this relationship for the average value of $\beta_2$ from three mass models as a function of $\nu$ and $\zeta$. Generally, high values of $\beta_2$ occur near the origin, which corresponds to being midshell for both protons and neutrons.

## VI. RESULTS AND DISCUSSION

It is possible that the most effective machine learning model can be trained using less than the maximum number of physical features. This could occur if the impact of a physical feature has already been accounted for or if that feature is simply superfluous. To test for this, we began by initially including all possible physical features for each pair of mass model and machine learning approach. Then we implemented a Shapley value analysis to determine the hierarchy of the physical features used in each model. Shapley values are a concept of cooperative game theory that explains the contribution of each participant to the outcome [35]. In our implementation, the Shapley value analysis allowed us to test models with the lower contributing features removed.

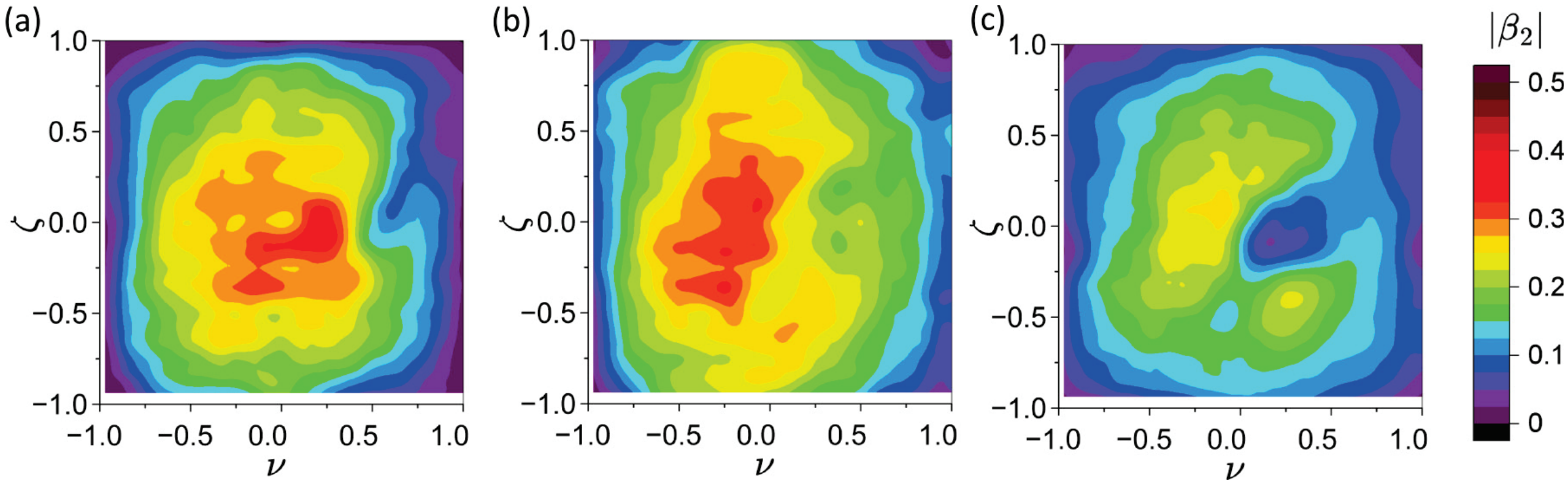

FIG. 4. The average $|\beta_2|$ deformation as a function of $\nu$ and $\zeta$ for (a) FRDM, (b) HFB, and (c) WS.

Interestingly, during this preliminary Shapley value analysis, it was observed that the ML approach, and not the particular mass model, consistently dictated which physical features were more or less influential.

For each mass model, we trained models using 13 groups of physical features; included in this was a two-feature group consisting of just $N$ and $Z$, which was intended to serve as a baseline. Table I summarizes the seven combinations of physical features that were ultimately found to result in the best fit. Feature Groups 5, 6, and 7 contain the full listings for the respective mass models. The features present in every best fit are $N$, $Z$, $T_Z$, $A$, $\nu$, $\zeta$, $N_E$, and $Z_E$.

Only Feature Group 2 did not include any shape-based features. In this case, the removal of all shape features resulted in a slight improvement for the WSRBFGPR model. In this case it is possible that the $\nu$ and $\zeta$ values, shown to have a correlation with $|\beta_2|$ in Fig. 4(c), were able to provide the information needed for the model.

TABLE I. Physical features used with ML models.

| Feature Group | Physical Features |
|---|---|
| 1 | $N, Z, T_Z, A, \nu, \zeta, N_E, Z_E, \beta_2$ |
| 2 | $N, Z, T_Z, A, \nu, \zeta, N_S, Z_S, N_E, Z_E$ |
| 3 | $N, Z, T_Z, A, \nu, \zeta, N_S, Z_S, N_E, Z_E, \beta_2$ |
| 4 | $N, Z, T_Z, A, \nu, \zeta, N_E, Z_E, \beta_2, \beta_4, R_C$ |
| 5 | $N, Z, T_Z, A, \nu, \zeta, N_S, Z_S, N_E, Z_E, \beta_2, \beta_4, R_C$ |
| 6 | $N, Z, T_Z, A, \nu, \zeta, N_S, Z_S, N_E, Z_E, \beta_2, \beta_4, \beta_6$ |
| 7 | $N, Z, T_Z, A, \nu, \zeta, N_S, Z_S, N_E, Z_E, \beta_2, \beta_3, \beta_4, \beta_6$ |

Table II contains standard deviations ($\sigma$) and $\overline{AE}$ for the original models and the new ML models. The subscript 12 corresponds to the AME 2012 dataset, and the subscript 20 corresponds to AME 2020 data. The LSBET and GPR models generally outperform SVM and FCNN.

Generally speaking, our approach did not improve the WSRBF model as much as the other mass models. The Radial Basis Function correction had already improved the model, and further ML modeling appears to have led to overfitting. In particular, the WSRBFGPR model provides a clear demonstration of model overfitting. The $\sigma$ and $\overline{AE}$ values for the training set are on the eV scale with values of 31.2 and 23.9 eV, respectively. One might naively believe this to be the best binding energy model, but these values are in fact 6318 and 4965 times larger, respectively, in the test set. In other words, this model nearly exactly reproduces the training data, but it does far less well with new data. This case highlights the need for independent test sets.

Based on the results in Table II, one could argue that every model shown exhibits overfitting because when comparing the training and test sets, the test sets always perform less well. For example, the $\overline{AE}$ values for LSBET models are 6 to 13 times larger for the test set than in the training set. This may not be entirely unexpected considering the nature of the data in the two sets. The experimental uncertainties are more than twice as large in the test set as in the training set. One can similarly compare how well the original models perform in the same comparison. The original mass models consistently perform better when compared to the training set than they do when compared to the test set. In general, the process of using the minimum $\overline{AE}_{20\mathrm{Test}}$ values to determine the best model allows one to find the models with the greatest potential to interpolate and extrapolate to nearby values.

One virtue of the best LSBET models is that they often require fewer physical features than the other ML approaches. Additionally, regarding modeling binding energy residuals, they often outperform the other approaches regarding evaluation metrics from the test set. The LSBET models take advantage of the known benefit of random forest approaches, which is the diminishing effect that each subsequent learner has on the models. This results in these models not being inherently overfitting, as the process obeys the Law of Large

TABLE II. Best trained models and corresponding evaluation metrics for $\Delta B_{\mathrm{FRDM}}$, $\Delta B_{\mathrm{HFB}}$, $\Delta B_{\mathrm{WS}}$, and $\Delta B_{\mathrm{WSRBF}}$ using both AME 2012 [17] and AME 2020 [18] data.

| Model Name | Feature Group | $\sigma_{12\mathrm{Train}}$ (MeV) | $\overline{AE}_{12\mathrm{Train}}$ (MeV) | $\sigma_{12}$ (MeV) | $\overline{AE}_{12}$ (MeV) | $\sigma_{20\mathrm{Test}}$ (MeV) | $\overline{AE}_{20\mathrm{Test}}$ (MeV) | $\sigma_{20}$ (MeV) | $\overline{AE}_{20}$ (MeV) |
|---|---|---|---|---|---|---|---|---|---|
| FRDM [20] | | 0.571 | 0.402 | 0.579 | 0.410 | 0.727 | 0.496 | 0.606 | 0.422 |
| FRDMSVM | 7 | 0.235 | 0.138 | 0.254 | 0.150 | 0.422 | 0.240 | 0.284 | 0.159 |
| FRDMGPR | 7 | 0.067 | 0.044 | 0.118 | 0.063 | 0.259 | 0.165 | 0.133 | 0.070 |
| FRDMFCNN | 3 | 0.111 | 0.084 | 0.153 | 0.100 | 0.337 | 0.189 | 0.182 | 0.105 |
| FRDMLSBET | 3 | 0.017 | 0.013 | 0.101 | 0.037 | 0.266 | 0.164 | 0.122 | 0.046 |
| HFB [21] | | 0.557 | 0.425 | 0.570 | 0.434 | 0.693 | 0.514 | 0.587 | 0.443 |
| HFBSVM | 5 | 0.322 | 0.209 | 0.339 | 0.221 | 0.482 | 0.313 | 0.360 | 0.230 |
| HFBGPR | 5 | 0.161 | 0.113 | 0.204 | 0.132 | 0.404 | 0.262 | 0.233 | 0.144 |
| HFBFCNN | 5 | 0.241 | 0.177 | 0.267 | 0.192 | 0.441 | 0.303 | 0.293 | 0.203 |
| HFBLSBET | 4 | 0.055 | 0.042 | 0.148 | 0.072 | 0.378 | 0.247 | 0.179 | 0.085 |
| WS [22] | | 0.286 | 0.226 | 0.298 | 0.233 | 0.327 | 0.253 | 0.295 | 0.231 |
| WSSVM | 6 | 0.177 | 0.124 | 0.196 | 0.135 | 0.249 | 0.178 | 0.194 | 0.135 |
| WSGPR | 6 | 0.046 | 0.032 | 0.089 | 0.048 | 0.185 | 0.129 | 0.094 | 0.053 |
| WSFCNN | 6 | 0.111 | 0.085 | 0.150 | 0.101 | 0.228 | 0.161 | 0.144 | 0.100 |
| WSLSBET | 3 | 0.021 | 0.016 | 0.094 | 0.038 | 0.181 | 0.128 | 0.085 | 0.041 |
| WSRBF [22,25] | | 0.168 | 0.131 | 0.170 | 0.132 | 0.253 | 0.178 | 0.189 | 0.141 |
| WSRBFSVM | 6 | 0.070 | 0.037 | 0.082 | 0.046 | 0.214 | 0.133 | 0.116 | 0.058 |
| WSRBFGPR | 2 | 0.000 | 0.000 | 0.049 | 0.014 | 0.197 | 0.119 | 0.090 | 0.029 |
| WSRBFFCNN | 6 | 0.085 | 0.062 | 0.094 | 0.067 | 0.199 | 0.126 | 0.118 | 0.075 |
| WSRBFLSBET | 1 | 0.023 | 0.017 | 0.059 | 0.031 | 0.189 | 0.119 | 0.088 | 0.039 |
| FMTE | | 0.015 | 0.012 | 0.081 | 0.031 | 0.164 | 0.112 | 0.076 | 0.034 |

Numbers [31], due to the consequence of the number of learners.

Comparisons of the WS with ML models against the original WSRBF model open a window to comparing results from ML approaches against the Radial Basis Function correction. Three of the four ML approaches (WSGPR, WSFCNN, and WSLSBET) outperform the corresponding WSRBF model when it comes to the test data and full AME 2020.

Figure 5 shows the $\Delta B$ values for each initial model, as well as the corresponding residual predicted by ML. This demonstrates that for the SVM-[Figs. 5(a)–5(d)] and GPR-based [Figs. 5(e)–5(h)] models the residuals predicted are localized, meaning that they do not extend in regions where the data are not known, specifically, toward the neutron drip line. The SVM models with low $C$ values result in smoother potentials. For GPR models, the $\sigma_l$ values set the maximum possible model value. The $\sigma_l$ values are lower for the WS and WSRBF residuals which were smaller on average. Additionally, it is worth noting that both the WSGPR and WSRBFGPR contain a linear basis function which provides a gradual adjustment of the mean value. In the case of these models, this effectively provided a tilt to the underlying background, which can be seen as the green color that appears for the high $N$ values in Fig. 5(g).

In regions further from stability, the FCNN models shown in Figs. 5(i)–5(l) occasionally predict residuals that are off scale compared to the original models shown in Fig. 2 while the LSBET models [Figs. 5(m)–5(p)] predict residuals that are comparable to the experimental $\Delta B$ values, even in regions far from stability. This further justifies the case that the LSBET is the preferable approach for modeling binding energy residuals.

### A. Understanding the models using shapley values

The Shapley values shown in Fig. 6 have been sorted in order of overall importance to the model. For example, the $N_E$ and $Z_E$ values play a highly impactful role (both were in the top three) for each of the WSRBF-based models, shown in the last column of Fig. 6 [see Figs. 6(d), 6(h), 6(l) and 6(p)]. In comparison, for the WSRBF-based models, the higher-order deformation parameters ($\beta_4$ and $\beta_6$) were either among the four least impactful of the models or they were not used.

Comparison of Shapley values can provide insight into the dominant features and the overall complexity of each model. The features that have a color gradient from left to right in Fig. 6 are best described as being either monotonically increasing or decreasing. This corresponds to high (and low) values that impact the model consistently in one manner or the other. The best example of this is the dependence on the mass number, as shown in Fig. 6(g), which corresponds to the WSGPR model. The high values of $A$ indicate negative Shapley values and the low values of $A$ indicate positive Shapley values.

The impact of WSRBF having already corrected the model with respect to measured masses means that the normally dominant features $A$, $N$, $Z$, and $T_Z$ in each of the SVM models and most of the GPR models seen in the first three columns do not persist in the WSRBF model.

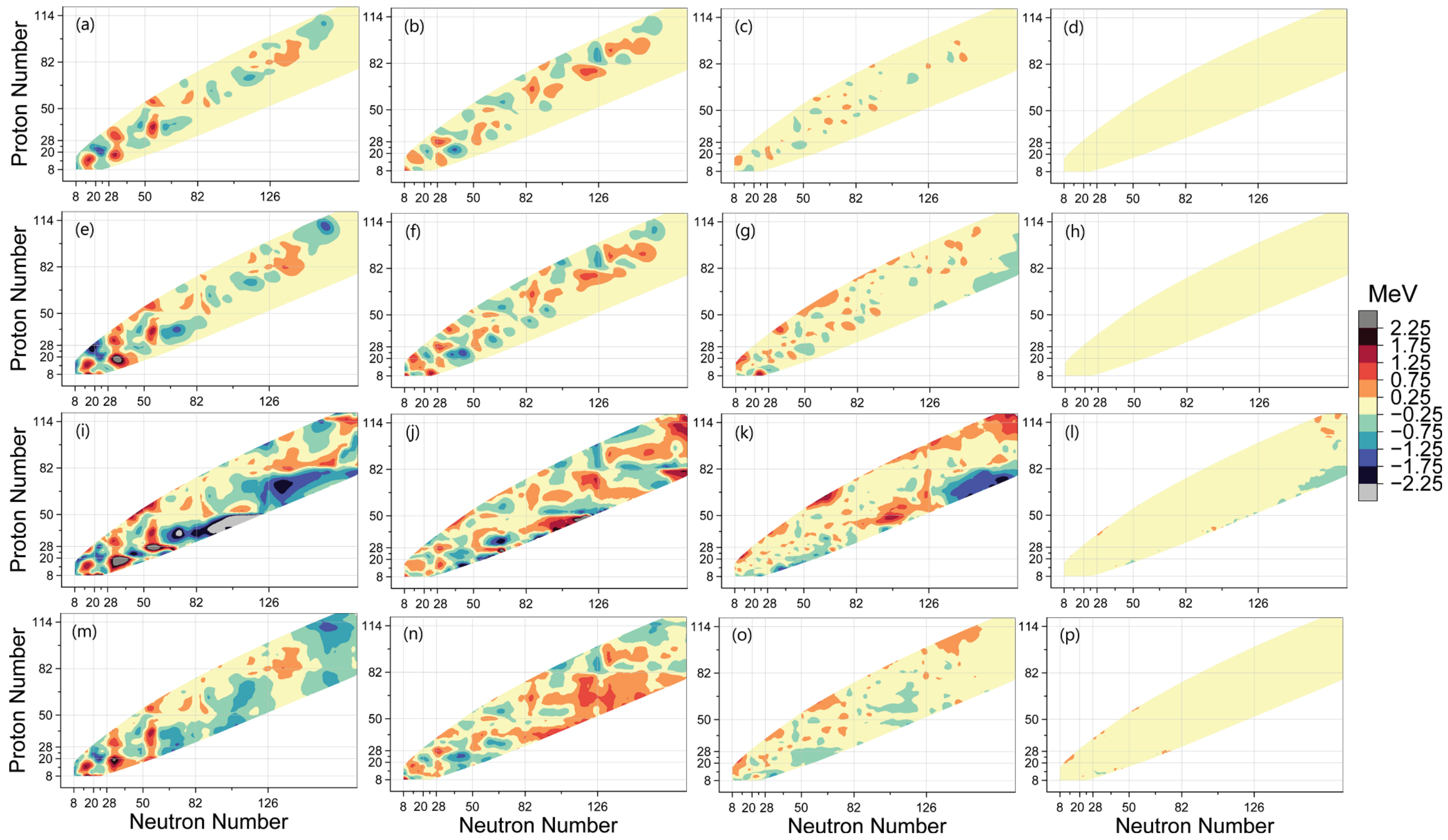

FIG. 5. Machine learning models arranged for each ML approach by row and for each mass model by column. The models are (a) FRDMSVM with $C = 0.4409$ and $\epsilon = 0.0441$; (b) HFBSVM with $C = 0.5076$ and $\epsilon = 0.0508$; (c) WSSVM with $C = 0.2794$ and $\epsilon = 0.0279$; (d) WSRBFSVM with $C = 0.1552$ and $\epsilon = 0.0155$; (e) FRDMGPR with zero basis function, $\sigma_l = 6.84$, and $\sigma_f = 0.404$; (f) HFBGPR with zero basis function, $\sigma_l = 6.05$, and $\sigma_f = 1.03$; (g) WSGPR with linear basis function, $\sigma_l = 4.93$, and $\sigma_f = 0.545$; (h) WSRBFGPR with linear basis function, $\sigma_l = 1.88$, and $\sigma_f = 0.227$; (i) FRDMFCNN with $\lambda = 6.25 \times 10^{-4}$; (j) HFBFCNN with $\lambda = 1.84 \times 10^{-3}$; (k) WSFCNN with $\lambda = 6.04 \times 10^{-4}$; (l) WSRBFFCNN with $\lambda = 3.88 \times 10^{-4}$; (m) FRDMLSBET with a minimum leaf size of 20 and $\eta = 0.212$; (n) HFBLSBET with a minimum leaf size of 25 and $\eta = 0.111$; (o) WSLSBET with a minimum leaf size of 28 and $\eta = 0.159$; and (p) WSRBFLSBET with a minimum leaf size of 31 and $\eta = 0.131$.

LSBET models, excluding WSRBFLSBET, are generally not very sensitive to the $N_E$ and $Z_E$, or $N_S$ and $Z_S$, values. Generally, for FRDMLSBET, HFBLSBET, and WSLSBET the $\beta_2$ value is of midlevel importance.

### B. Generation of a composite mass model

The goal of this work has not been to produce a large number of decent models but instead to generate one superior mass model that can be used to estimate binding energies for experimental measurements, astrophysical calculations, and to explore the limits of nuclear matter. For this reason, we have developed a weighted ensemble that used some of the models generated here as well as one model from our related work [36].

The ensemble of various models can increase the overall accuracy of predictions for several reasons. First, individual models can be weak learners in the entire domain or parts of the domain. The errors made by individual models can be mitigated by ensembling weak learners, especially when the errors are of a statistical nature, stemming from limited sample size of the training data, as we have in our case. Therefore, it is not unreasonable to expect that ensembling a set of the best models will, in fact, produce a model that performs better than any of the individual models. Although the simplest way to ensemble a regression model is to take the average of their predictions, this might not provide an optimal ensemble. Our approach will include testing of both equal and varying ratios of models.

The binding energy for the ensembled model resulted from the sum of weighted models:

$$B_{\text{ens}}(N, Z) = \sum_{i=1}^{M} w_i B_i(N, Z), \tag{16}$$

where $M$ is the number of models included, $w_i$ is the weight for the $i$th model, and $B_i(N, Z)$ is the corresponding binding energy for a specific isotope in that model. Individual weights were determined using amplitudes, denoted as $a_i$, that were cycled through using an integer counter. For example, the weight of model number 1 is determined using

$$w_1 = a_1^2 \Big/ \sum_{i=1}^{M} a_i^2, \tag{17}$$

where the denominator normalizes the weights so they satisfy the condition

$$\sum_{i=1}^{M} w_i = 1. \tag{18}$$

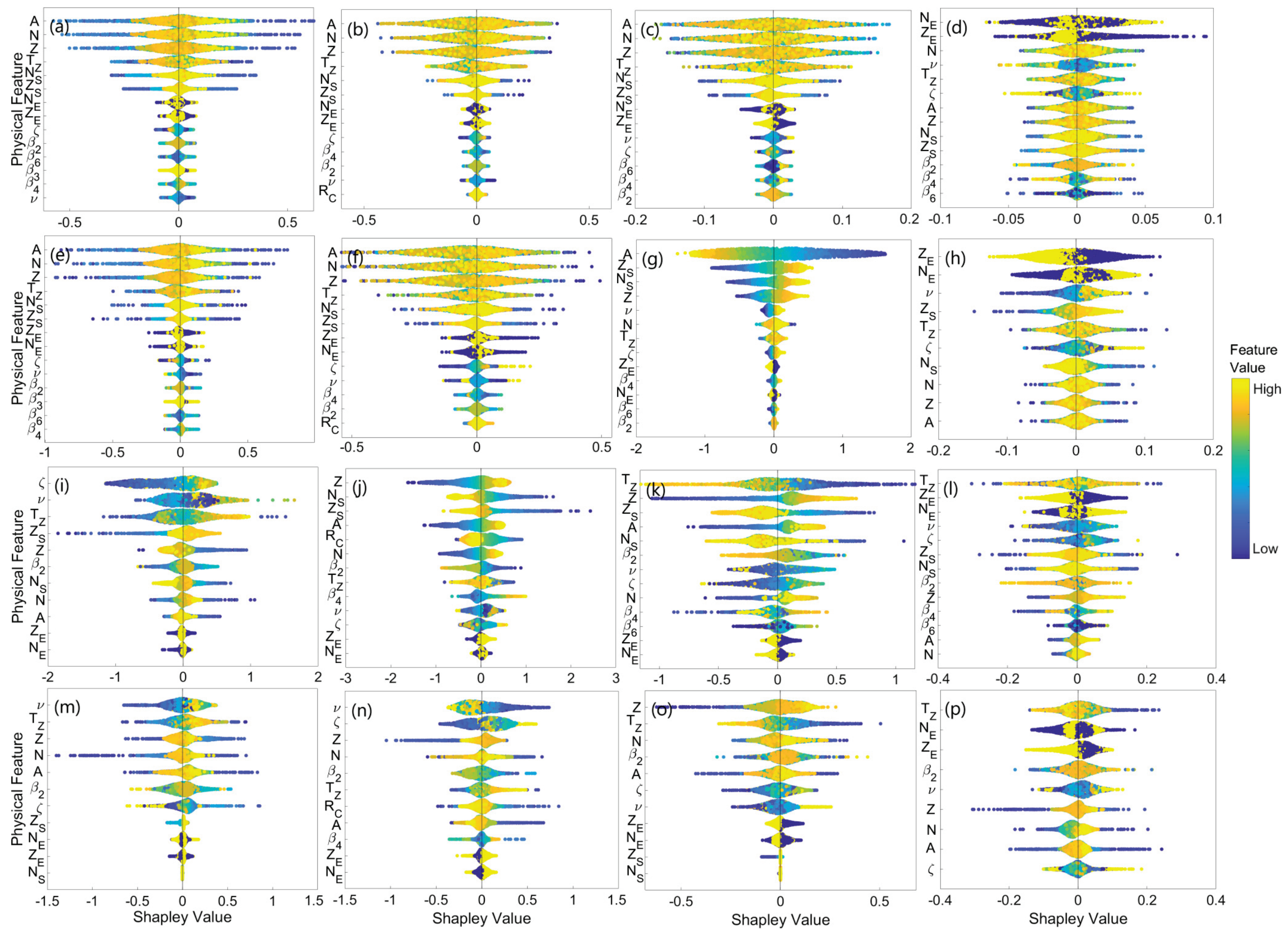

FIG. 6. Distribution of local Shapley values for the 16 best models arranged for each ML approach by row and for each mass model by column. These correspond to (a) FRDMSVM, (b) HFBSVM, (c) WSSVM, (d) WSRBFSVM, (e) FRDMGPR, (f) HFBGPR, (g) WSGPR, (h) WSRBFGPR, (i) FRDMFCNN, (j) HFBFCNN, (k) WSFCNN, (l) WSRBFFCNN, (m) FRDMLSBET, (n) HFBLSBET, (o) WSLSBET, and (p) WSRBFLSBET. The vertical spread in points represents how many values are located in the same region. The predictor value color demonstrates if the value for the given input predictor was high or low.

We tested a variety of combinations involving as many as ten models from this work and Ref. [16]. Ultimately, the LSBET models were selected because, first and foremost, they interpolate well. Additionally, in Ref. [16], we found that the best LSBET models reproduced a critical feature at $N = Z$ that emerged in Garvey-Kelson relations whether or not the original mass model contained that structure. Additionally, the LSBET approach consistently results in extrapolated values on scale with what is seen in the experimental values near stability. The WSRBFLSBET model was omitted because it is essentially the same as the WSLSBET.

The lowest $\overline{AE}$ of the test set was also used to determine the composite model. The two liquid-drop-based LSBET models from Ref. [16] were not used, because it was found that the composite model works best when they had amplitudes equal to zero, meaning that including them would not provide an overall benefit. However, the DZ-based LSBET model, DZLSBET from Ref. [16], was found to provide an improvement. An optimization with amplitude values ranging from 0 to 60 determined that the four LSBET models performed best when combined with the following amplitudes: $a_{\mathrm{WSLSBET}} = 55$, $a_{\mathrm{DZLSBET}} = 51$, $a_{\mathrm{FRDMLSBET}} = 19$, and $a_{\mathrm{HFBLSBET}} = 14$, Using Eq. (17). These amplitudes correspond to the model being comprised of 48.9% WSLSBET, 42.1% DZLSBET, 5.8% FRDMLSBET, and 3.2% HFBLSBET. This model has been named the Four Model Tree Ensemble (FMTE).

Modifications of these weights by a percent or two will result in different binding energy values for the composite model, particularly far from stability, but changes of that magnitude were observed to have less than a 1 keV effect on the comparison metrics described in Tables II and III. Regarding the alternate option of equally weighing all four models, on average using the FMTE's ratios resulted in a 16% improvement for the 16 metrics shown in Tables II and III, when compared to using an equal weight of 25% for each.

Figure 7 contains the $\Delta B$ values for the FMTE model. Further from stability are regions that exceed $|\Delta B| = 250$ keV, but the majority of values have $|\Delta B| < 50$ keV.

TABLE III. Model evaluation metrics from recent mass measurements from Refs. [42–66] for five original mass models, the corresponding LSBET models, and the FMTE model.

| Model Name | $\sigma_{\mathrm{Recent}}$ (MeV) | $\overline{AE}_{\mathrm{Recent}}$ (MeV) | $\sigma_{\mathrm{inTrain}}$ (MeV) | $\overline{AE}_{\mathrm{inTrain}}$ (MeV) | $\sigma_{\mathrm{inAME}}$ (MeV) | $\overline{AE}_{\mathrm{inAME}}$ (MeV) | $\sigma_{\mathrm{New}}$ (MeV) | $\overline{AE}_{\mathrm{New}}$ (MeV) |
|---|---|---|---|---|---|---|---|---|
| DZ [19] | 0.570 | 0.398 | 0.411 | 0.315 | 0.693 | 0.486 | 0.674 | 0.449 |
| FRDM [20] | 0.836 | 0.631 | 0.724 | 0.558 | 0.945 | 0.701 | 0.743 | 0.549 |
| HFB [21] | 0.647 | 0.484 | 0.578 | 0.423 | 0.668 | 0.497 | 0.801 | 0.614 |
| WS [22] | 0.341 | 0.267 | 0.299 | 0.243 | 0.318 | 0.259 | 0.488 | 0.360 |
| WSRBF [22,25] | 0.267 | 0.186 | 0.196 | 0.156 | 0.241 | 0.178 | 0.445 | 0.295 |
| DZLSBET [16] | 0.207 | 0.115 | 0.058 | 0.038 | 0.206 | 0.157 | 0.417 | 0.258 |
| FRDMLSBET | 0.246 | 0.133 | 0.060 | 0.040 | 0.300 | 0.199 | 0.420 | 0.261 |
| HFBLSBET | 0.371 | 0.209 | 0.077 | 0.057 | 0.362 | 0.281 | 0.761 | 0.533 |
| WSLSBET | 0.180 | 0.099 | 0.060 | 0.041 | 0.150 | 0.117 | 0.379 | 0.228 |
| WSRBFLSBET | 0.193 | 0.102 | 0.060 | 0.040 | 0.186 | 0.127 | 0.396 | 0.253 |
| FMTE | 0.175 | 0.090 | 0.058 | 0.038 | 0.142 | 0.111 | 0.376 | 0.206 |

### C. Garvey-Kelson relations

The Garvey-Kelson mass relations provide a means of testing the relative output of a binding energy model by comparing nearby values [37]. The two relationships, Eqs. (1) and (3) from Ref. [38], used are

$$
\begin{aligned}
&M(N+2, Z-2) - M(N, Z) \\
&\quad + M(N, Z-1) - M(N+1, Z-2) \\
&\quad + M(N+1, Z) - M(N+2, Z-1) \approx 0,
\end{aligned} \tag{19}
$$

for $N \geqslant Z$, and

$$
\begin{aligned}
&M(N-2, Z+2) - M(N, Z) \\
&\quad + M(N-1, Z) - M(N-2, Z+1) \\
&\quad + M(N, Z+1) - M(N-1, Z+2) \approx 0,
\end{aligned} \tag{20}
$$

for $N < Z$, where $M(N, Z)$ is the mass of an isotope with the corresponding number of protons and neutrons.

Figure 8 demonstrates the values of the Garvey-Kelson relations for both the experimental measurements and the FMTE model. The FMTE model reproduces the Wigner cusp at $N = Z$ (see, e.g., Refs. [39–41]), where these relations are known to deviate from zero, as discussed by Garvey *et al.* in Ref. [37]. The FMTE model predicts that this cusp phenomenon continues for nuclei near the $N = Z$ line for proton-rich nuclei beyond what has been measured to date. Elsewhere, the FMTE model is smooth, resulting in Garvey-Kelson relation values that are nearly zero in a comparable manner to what is seen in the experimental values.

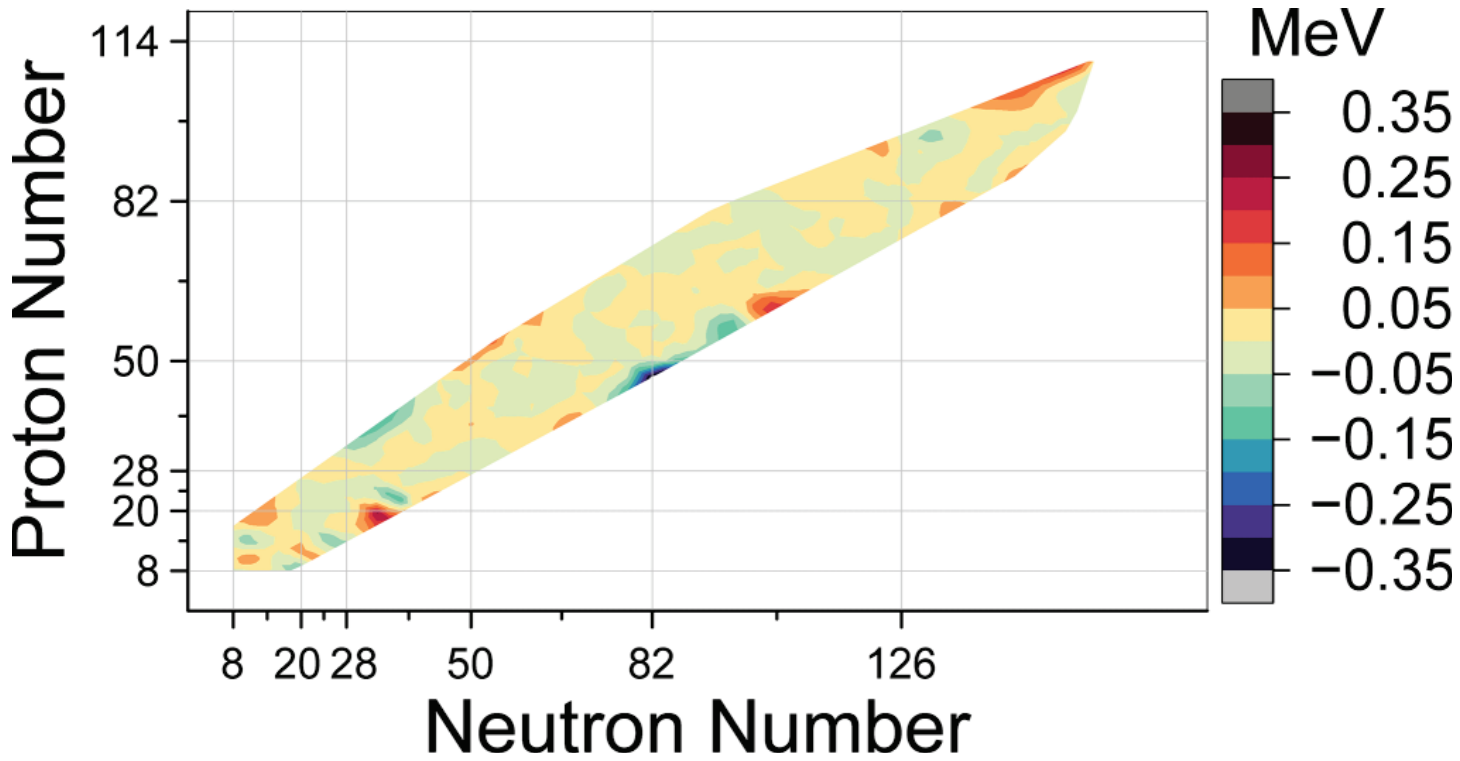

FIG. 7. Binding energy difference $\Delta B$ between the FMTE and AME 2020 [18]. Note that the absolute range displayed here is approximately one-sixth of that shown in Figs. 2 and 5.

### D. Comparisons using recent mass measurements

An additional test of the results from FMTE and the constituent LSBET models was conducted using a survey of recent (i.e., post–AME 2020) mass measurements. This survey identified 207 new mass measurements from Refs. [42–66]. Of these recent measurements, 106 were for isotopes used in the training set. Among these measurements, the values changed by 36 keV on average and the average experimental uncertainty is 25 keV. Another 68 measurements were of isotopes included in either AME 2012 or AME 2020. For these 68 isotopes, the average change was 172 and 48 keV for these two groups, respectively. The average experimental uncertainty for these 68 measurements was 27 keV. There were an additional 33 measurements for isotopes not included in AME 2012 or AME 2020. The average experimental uncertainty for these newly measured isotopes is 132 keV.

Figure 9 demonstrates the range of isotopes with $N > 7$ and $Z > 7$ in the AME 2012 and AME 2020, as well as the location of these three subgroups of recent measurements. Many of the new measurements are near regions where the FMTE performs less well in Fig. 7 (e.g., near $Z = 40$, and $N = 66$, where FMTE deviates from experimental values by about 150 keV).

Table III contains the comparisons for all recent measurements, recent remeasurements of isotopes used in training, recent remeasurements of isotopes included in either AME 2012 or AME 2020, and newly measured isotopes. When comparing the performance of remeasurements for isotopes in the training set with newly measured isotopes, evidence of overfitting exists for the FMTE and the LSBET models because the new measurements do not perform as well in the latter. A decrease in performance, from comparing $\overline{AE}_{\mathrm{inAME}}$ to $\overline{AE}_{\mathrm{New}}$ also occurs in the original models with one exception, where FRDM improved.

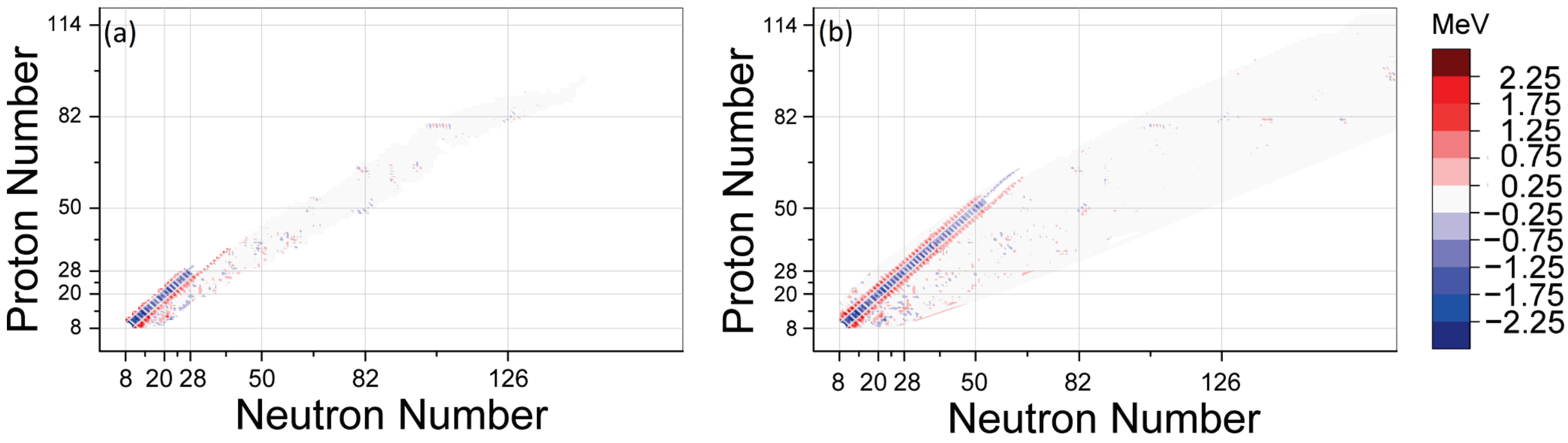

FIG. 8. The Garvey-Kelson mass relationships for (a) experimental values from AME 2020 [18] and (b) FMTE.

An alternative comparison is of each LSBET model with its corresponding original mass model. In each case, the LSBET model performs better across all metrics. For the $\overline{AE}_{\mathrm{New}}$ values, the average drops from 454 keV for the five original models to 307 keV for the five LSBET models and to 206 keV for the FMTE.

### E. Extrapolation

Part of the purpose in producing the FMTE model is to have a model that can be useful for astrophysical calculations. Here we will focus only on the LSBET models for neutron-rich nuclei where the models can deviate substantially, as demonstrated in Fig. 1.

Figure 10 shows the four original mass models, and the corresponding four LSBET-based residual models used to create FMTE, along with the FMTE model for six isotopic chains. Figure 10 demonstrates the general characteristic of the LSBET models, which is that they generally correct each model inward toward a midpoint. The exception to that observation is the HFBLSBET, which is a correction that on occasion adds highest-valued models, which can be seen as the dotted line for HFBLSBET being above the solid line for HFB.

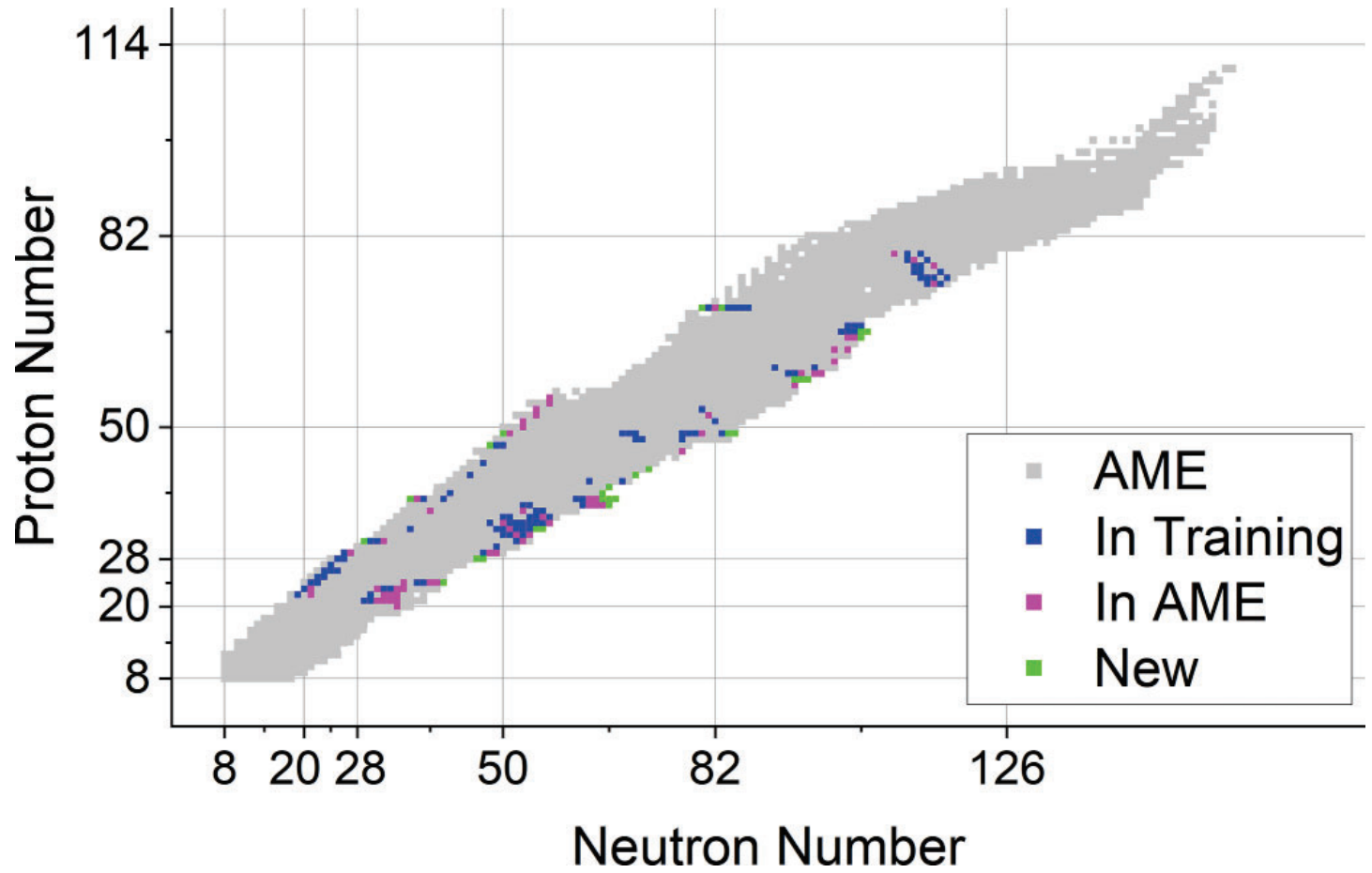

FIG. 9. Recent mass measurements grouped by isotopes that were included in the training set, isotopes found in either AME 2012 or AME 2020, and new isotopes not included in either AME set. All other isotopes found in either AME 2012 or AME 2020 are indicated in gray.

In general, Fig. 10 demonstrates how the FMTE model behaves. It is the optimized weighted average of the LSBET models as it consists of models with minor improvements on four of the most commonly used mass models.

## VII. SUMMARY AND CONCLUSION

In this paper, we have further explored using machine learning as a means of modeling binding energy residuals. Here, we have used contemporary mass models that contain shape features to determine the residuals. In all but one case, we have found that the inclusion of shape features resulted in a better-performing model than omitting those features would have.

Using $\overline{AE}_{20\mathrm{Test}}$ as a success metric, the WSLSBET performed well and, in general, the LSBET technique was successful. In addition to performing well regarding statistical metrics, the LSBET models have extrapolations that are comparable with the original values, as shown in Fig. 5. The best LSBET models also generally require the fewest physical features.

This work culminates in the FMTE model which primarily combines the WSLSBET and DZLSBET (from Ref. [16]), and to a lesser extent the FRDMLSBET and HFBLSBET models. When comparing the models discussed in this paper, the FMTE appears to interpolate and extrapolate better than any other model, as demonstrated in the evaluation metrics involving the test set and the AME 2020 data included in Table II. Furthermore, in the compilation of recent mass measurements included in Table III, the LSBET models consistently outperform the values $\overline{AE}_{\mathrm{New}}$ for the original mass models by on average about 150 keV, and the FMTE outperforms those original models by roughly 250 keV.

The $\overline{AE}$ value of FMTE for the $N > 7$ and $Z > 7$ isotopes in AME 2020 is 34 keV, which is comparable to the average experimental uncertainty of 23 keV for the AME 2020. The corresponding standard deviation value is only 76 keV. However, the standard deviation was 376 keV for a set of 33 new mass measurements for previously unmeasured isotopes. Based on these data, the FMTE falls short of the desired 50 keV target needed for enhanced understanding of astrophysics from Clark *et al.* [8]. In addition, we have seen likely evidence that overfitting of the models that were used in FMTE may have occurred.

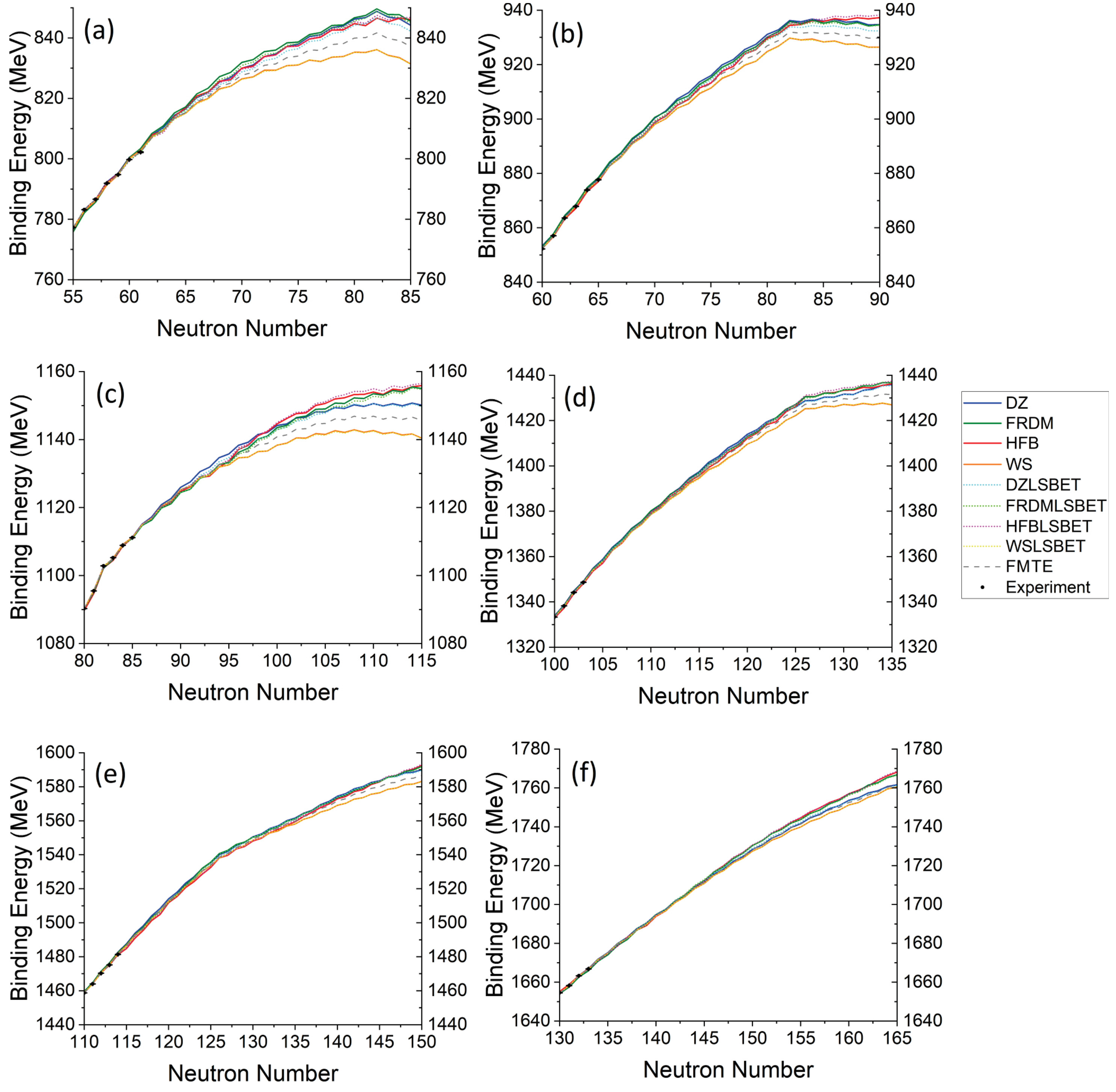

FIG. 10. Mass model extrapolation comparison for neutron-rich (a) krypton, (b) zirconium, (c) tin, (d) gadolinium, (e) hafnium, and (f) lead isotopes. Experimental values from AME 2020 [18] are included as black circles. Solid lines indicate the four original models from Refs. [19–21], and [22]. Dotted lines denote LSBET models including DZLSBET from Ref. [16]. The FMTE model is included as the gray dashed lines.

Regarding overfitting of LSBET-based models, if the original mass models are uniformly (not regionally) accurate and the ML models are trained on a truly random and representative set, then the LSBET models trained on a sufficient quantity of data will not overfit, even if an excessive number of learners is used. A fundamental assumption in this work was that the training set was a valid representation of the data elsewhere. It should be noted that this may not be the case. Shell structure is known to evolve, as thoroughly discussed by Otsuka *et al.* [67]. It is entirely possible that every mass model, including the original mass models, is inaccurate in regions far from stability because it lacks the necessary physics as input. For this reason, new experimental measurements are critical for further refinement of mass models.

Nevertheless, on the basis of all of the comparisons demonstrated, the FMTE provides the best interpolation and near extrapolations, making it valuable for new experimental measurements. Additionally, in regions of astrophysical interest, far from stability, the FMTE acts as an improved weighted average of models.

In the future, we intend to use the FMTE binding energies as input features for future ML calculations. Possible physical properties for new models include low-lying states and transition probabilities, charge radii, and half-life values.

## DATA AVAILABILITY

The data that support the findings of this article are openly available [68], embargo periods may apply.

---

[1] A. Sobiczewski, Y. Litvinov, and M. Palczewski, Detailed illustration of the accuracy of currently used nuclear-mass models, At. Data Nucl. Data Tables **119**, 1 (2018).

[2] T. Y. Hirsh, N. Paul, M. Burkey, A. Aprahamian, F. Buchinger, S. Caldwell, J. A. Clark, A. F. Levand, L. L. Ying, S. T. Marley, G. E. Morgan, A. Nystrom, R. Orford, A. P. Galván, J. Rohrer, G. Savard, K. S. Sharma, and K. Siegl, First operation and mass separation with the caribu mr-tof, Nucl. Instrum. Methods Phys. Res. Sect. B 376, 229 (2016), proceedings of the XVIIth International Conference on Electromagnetic Isotope Separators and Related Topics.

[3] J. Wei, C. Alleman, H. Ao, B. Arend, D. Barofsky, S. Beher, G. Bollen, N. Bultman, F. Casagrande, W. Chang, Y. Choi, S. Cogan, P. Cole, C. Compton, M. Cortesi, J. Curtin, K. Davidson, S. D. Carlo, X. Du, K. Elliott *et al.*, Technological developments and accelerator improvements for the FRIB beam power ramp-up, J. Instrum. **19**, T05011.

[4] M. Vilen, J. M. Kelly, A. Kankainen, M. Brodeur, A. Aprahamian, L. Canete, T. Eronen, A. Jokinen, T. Kuta, I. D. Moore, M. R. Mumpower, D. A. Nesterenko, H. Penttilä, I. Pohjalainen, W. S. Porter, S. Rinta-Antila, R. Surman, A. Voss, and J. Äystö, Precision mass measurements on neutron-rich rare-earth isotopes at JYFLTRAP: Reduced neutron pairing and implications for *r*-process calculations, Phys. Rev. Lett. **120**, 262701 (2018).

[5] S. Brett, I. Bentley, N. Paul, R. Surman, and A. Aprahamian, Sensitivity of the *r*-process to nuclear masses, Eur. Phys. J. A **48**, 184 (2012).

[6] A. Aprahamian, I. Bentley, M. Mumpower, and R. Surman, Sensitivity studies for the main *r* process: Nuclear masses, AIP Adv. **4**, 041101 (2014).

[7] D. Martin, A. Arcones, W. Nazarewicz, and E. Olsen, Impact of nuclear mass uncertainties on the *r* process, Phys. Rev. Lett. **116**, 121101 (2016).

[8] J. Clark, G. Savard, M. Mumpower, and A. Kankainen, Precise mass measurements of radioactive nuclides for astrophysics, Eur. Phys. J. A **59**, (2023).

[9] M.-H. Chen, L.-X. Li, E.-W. Liang, and N. Wang, Impact of nuclear mass models on *r*-process nucleosynthesis and heavy element abundances in *r*-process-enhanced metal-poor stars, Astron. Astrophys. **693**, A1 (2025).

[10] A. E. Lovell, A. T. Mohan, T. M. Sprouse, and M. R. Mumpower, Nuclear masses learned from a probabilistic neural network, Phys. Rev. C **106**, 014305 (2022).

[11] E. Yüksel, D. Soydaner, and H. Bahtiyar, Nuclear mass predictions using machine learning models, Phys. Rev. C **109**, 064322 (2024).

[12] O. Kitouni, N. Nolte, S. Trifinopoulos, S. Kantamneni, and M. Williams, NuCLR: Nuclear Co-learned representations, arXiv:2306.06099.

[13] S. Choi, K. Kim, Z. He, Y. Kim, and T. Kajino, Deep learning for nuclear masses in deformed relativistic Hartree-Bogoliubov theory in continuum, arXiv:2411.19470.

[14] Y. Lu, T. Shang, P. Du, J. Li, H. Liang, and Z. Niu, Nuclear mass predictions based on a convolutional neural network, Phys. Rev. C **111**, 014325 (2025).

[15] G.-P. Liu, H.-L. Wang, Z.-Z. Zhang, and M.-L. Liu, Model-repair capabilities of tree-based machine-learning algorithms applied to theoretical nuclear mass models, Phys. Rev. C **111**, 024306 (2025).

[16] I. Bentley, J. Tedder, M. Gebran, and A. Paul, High precision binding energies from physics-informed machine learning, Phys. Rev. C **111**, 034305 (2025).

[17] M. Wang, G. Audi, A. Wapstra, F. Kondev, M. MacCormick, X. Xu, and B. Pfeiffer, The Ame2012 atomic mass evaluation, Chin. Phys. C **36**, 1603 (2012).

[18] M. Wang, W. Huang, F. Kondev, G. Audi, and S. Naimi, The ame 2020 atomic mass evaluation (ii). Tables, graphs and references, Chin. Phys. C **45**, 030003 (2021).

[19] J. Duflo and A. P. Zuker, Microscopic mass formulas, Phys. Rev. C **52**, R23 (1995).

[20] P. Möller, A. Sierk, T. Ichikawa, and H. Sagawa, Nuclear ground-state masses and deformations: FRDM(2012), At. Data Nucl. Data Tables **109-110**, 1 (2016).

[21] S. Goriely, N. Chamel, and J. M. Pearson, Further explorations of Skyrme-Hartree-Fock-Bogoliubov mass formulas. XVI. inclusion of self-energy effects in pairing, Phys. Rev. C **93**, 034337 (2016).

[22] N. Wang, M. Liu, X. Wu, and J. Meng, Surface diffuseness correction in global mass formula, Phys. Lett. B **734**, 215 (2014).

[23] https://groups.nscl.msu.edu/frib/rates/fribrates.html.

[24] M. Mumpower, R. Surman, G. McLaughlin, and A. Aprahamian, The impact of individual nuclear properties on *r*-process nucleosynthesis, Prog. Part. Nucl. Phys. **86**, 86 (2016).

[25] N. Wang and M. Liu, Nuclear mass predictions with a radial basis function approach, Phys. Rev. C **84**, 051303(R) (2011).

[26] H. Drucker, C. J. C. Burges, L. Kaufman, A. Smola, and V. Vapnik, Support vector regression machines, in *Advances in Neural Information Processing Systems*, edited by M. Mozer, M. Jordan, and T. Petsche (MIT Press, Cambridge, MA, 1996), Vol. 9.

[27] J. P. Janet and H. J. Kulik, *Machine Learning in Chemistry* (American Chemical Society, Washington, DC, 2020).

[28] C. Rasmussen and C. Williams, *Gaussian Processes for Machine Learning* (MIT Press, Cambridge, MA, 2005).

[29] I. Bentley and M. Gebran, *Neural Networks in the Physical Sciences* (American Chemical Society, Washington, DC, 2025).

[30] L. Berlyand and P.-E. Jabin, *Mathematics of Deep Learning: An Introduction* (De Gruyter, Berlin, 2023).

[31] L. Breiman, Random forests, Mach. Learn. **45**, 5 (2001).

[32] T. Hastie, R. Tibshirani, and J. Friedman, *The Elements of Statistical Learning: Data Mining, Inference, and Prediction*, Springer series in statistics (Springer, Berlin, 2009).

[33] I. Bentley, Particle-hole symmetry numbers for nuclei, Indian J. Phys. **90**, 1069 (2016).

[34] R. Firestone, *Table of Isotopes CD-ROM* (Wiley-Interscience, 1999).

[35] L. S. Shapley, Notes on the $n$-Person Game-II: The Value of an $n$-Person Game, (RAND Corporation, 1951).

[36] I. Bentley, Y. C. Rodríguez, S. Cunningham, and A. Aprahamian, Shell structure from nuclear observables, Phys. Rev. C **93**, 044337 (2016).

[37] G. T. Garvey, W. J. Gerace, R. L. Jaffe, I. Talmi, and I. Kelson, Set of nuclear-mass relations and a resultant mass table, Rev. Mod. Phys. **41**, S1 (1969).

[38] J. Barea, A. Frank, J. G. Hirsch, P. V. Isacker, S. Pittel, and V. Velázquez, Garvey-Kelson relations and the new nuclear mass tables, Phys. Rev. C **77** 041304(R) (2008).

[39] W. Satula, D. Dean, J. Gary, S. Mizutori, and W. Nazarewicz, On the origin of the Wigner energy, Phys. Lett. B **407**, 103 (1997).

[40] J. Jänecke, T. W. O'Donnell, and V. I. Goldanskii, Isospin inversion, $n$-$p$ interactions, and quartet structures in $N = Z$ nuclei, Phys. Rev. C **66**, 024327 (2002).

[41] I. Bentley and S. Frauendorf, Relation between Wigner energy and proton-neutron pairing, Phys. Rev. C **88**, 014322 (2013).

[42] R. Silwal, C. Andreoiu, B. Ashrafkhani, J. Bergmann, T. Brunner, J. Cardona, K. Dietrich, E. Dunling, G. Gwinner, Z. Hockenbery, J. Holt, C. Izzo, A. Jacobs, A. Javaji, B. Kootte, Y. Lan, D. Lunney, E. Lykiardopoulou, T. Miyagi, M. Mougeot *et al.*, Summit of the $n = 40$ island of inversion: Precision mass measurements and *ab initio* calculations of neutron-rich chromium isotopes, Phys. Lett. B **833**, 137288 (2022).

[43] S. Giraud, L. Canete, B. Bastin, A. Kankainen, A. Fantina, F. Gulminelli, P. Ascher, T. Eronen, V. Girard-Alcindor, A. Jokinen, A. Khanam, I. Moore, D. Nesterenko, F. de Oliveira Santos, H. Penttilä, C. Petrone, I. Pohjalainen, A. De Roubin, V. Rubchenya, M. Vilen, *et al.*, Mass measurements towards doubly magic $^{78}$Ni: Hydrodynamics versus nuclear mass contribution in core-collapse supernovae, Phys. Lett. B **833**, 137309 (2022).

[44] M. Hukkanen, W. Ryssens, P. Ascher, M. Bender, T. Eronen, S. Grévy, A. Kankainen, M. Stryjczyk, O. Beliuskina, Z. Ge, S. Geldhof, M. Gerbaux, W. Gins, A. Husson, D. Nesterenko, A. Raggio, M. Reponen, S. Rinta-Antila, J. Romero, A. de Roubin *et al.*, Precision mass measurements in the zirconium region pin down the mass surface across the neutron Midshell at $N = 66$, Phys. Lett. B **856**, 138916 (2024).

[45] A. Valverde, F. Kondev, B. Liu, D. Ray, M. Brodeur, D. Burdette, N. Callahan, A. Cannon, J. Clark, D. Hoff, R. Orford, W. Porter, G. Savard, K. Sharma, and L. Varriano, Precise mass measurements of $a = 133$ isobars with the Canadian penning trap: Resolving the $Q_{\beta-}$ anomaly at $^{33}$Te, Phys. Lett. B **858**, 139037 (2024).

[46] D. Puentes, G. Bollen, M. Brodeur, M. Eibach, K. Gulyuz, A. Hamaker, C. Izzo, S. M. Lenzi, M. MacCormick, M. Redshaw, R. Ringle, R. Sandler, S. Schwarz, P. Schury, N. A. Smirnova, J. Surbrook, A. A. Valverde, A. C. C. Villari, and I. T. Yandow, High-precision mass measurements of the isomeric and ground states of $^{44}$V: Improving constraints on the isobaric multiplet mass equation parameters of the $A = 44$, $0^+$ quintet, Phys. Rev. C **101**, 064309 (2020).

[47] C. Izzo, J. Bergmann, K. A. Dietrich, E. Dunling, D. Fusco, A. Jacobs, B. Kootte, G. Kripkó-Koncz, Y. Lan, E. Leistenschneider, E. M. Lykiardopoulou, I. Mukul, S. F. Paul, M. P. Reiter, J. L. Tracy, C. Andreoiu, T. Brunner, T. Dickel, J. Dilling, I. Dillmann *et al.*, Mass measurements of neutron-rich indium isotopes for $r$-process studies, Phys. Rev. C **103**, 025811 (2021).

[48] I. Mukul, C. Andreoiu, J. Bergmann, M. Brodeur, T. Brunner, K. A. Dietrich, T. Dickel, I. Dillmann, E. Dunling, D. Fusco, G. Gwinner, C. Izzo, A. Jacobs, B. Kootte, Y. Lan, E. Leistenschneider, E. M. Lykiardopoulou, S. F. Paul, M. P. Reiter, J. L. Tracy *et al.*, Examining the nuclear mass surface of Rb and Sr isotopes in the $A \approx 104$ region via precision mass measurements, Phys. Rev. C **103**, 044320 (2021).

[49] S. F. Paul, J. Bergmann, J. D. Cardona, K. A. Dietrich, E. Dunling, Z. Hockenbery, C. Hornung, C. Izzo, A. Jacobs, A. Javaji, B. Kootte, Y. Lan, E. Leistenschneider, E. M. Lykiardopoulou, I. Mukul, T. Murböck, W. S. Porter, R. Silwal, M. B. Smith, J. Ringuette *et al.*, Mass measurements of $^{60-63}$Ga reduce x-ray burst model uncertainties and extend the evaluated $T = 1$ isobaric multiplet mass equation, Phys. Rev. C **104**, 065803 (2021).

[50] R. Orford, N. Vassh, J. A. Clark, G. C. McLaughlin, M. R. Mumpower, D. Ray, G. Savard, R. Surman, F. Buchinger, D. P. Burdette, M. T. Burkey, D. A. Gorelov, J. W. Klimes, W. S. Porter, K. S. Sharma, A. A. Valverde, L. Varriano, and X. L. Yan, Searching for the origin of the rare-earth peak with precision mass measurements across Ce–Eu isotopic chains, Phys. Rev. C **105**, L052802 (2022).

[51] W. S. Porter, E. Dunling, E. Leistenschneider, J. Bergmann, G. Bollen, T. Dickel, K. A. Dietrich, A. Hamaker, Z. Hockenbery, C. Izzo, A. Jacobs, A. Javaji, B. Kootte, Y. Lan, I. Miskun, I. Mukul, T. Murböck, S. F. Paul, W. R. Plaß, D. Puentes *et al.*, Investigating nuclear structure near $N = 32$ and $N = 34$: precision mass measurements of neutron-rich Ca, Ti, and V isotopes, Phys. Rev. C **106**, 024312 (2022).

[52] Y. M. Xing, C. X. Yuan, M. Wang, Y. H. Zhang, X. H. Zhou, Y. A. Litvinov, K. Blaum, H. S. Xu, T. Bao, R. J. Chen, C. Y. Fu, B. S. Gao, W. W. Ge, J. J. He, W. J. Huang, T. Liao, J. G. Li, H. F. Li, S. Litvinov, S. Naimi *et al.*, Isochronous mass measurements of neutron-deficient nuclei from $^{112}$Sn projectile fragmentation, Phys. Rev. C **107**, 014304 (2023).

[53] A. Jaries, M. Stryjczyk, A. Kankainen, L. Al Ayoubi, O. Beliuskina, P. Delahaye, T. Eronen, M. Flayol, Z. Ge, W. Gins, M. Hukkanen, D. Kahl, S. Kujanpää, D. Kumar, I. D. Moore, M. Mougeot, D. A. Nesterenko, S. Nikas, H. Penttilä, D. Pitman-Weymouth *et al.*, High-precision penning-trap mass measurements of Cd and In isotopes at JYFLTRAP remove the fluctuations in the two-neutron separation energies, Phys. Rev. C **108**, 064302 (2023).

[54] W. Xian, S. Chen, S. Nikas, M. Rosenbusch, M. Wada, H. Ishiyama, D. Hou, S. Iimura, S. Nishimura, P. Schury, A. Takamine, S. Yan, F. Browne, P. Doornenbal, F. Flavigny, Y. Hirayama, Y. Ito, S. Kimura, T. M. Kojima, J. Lee *et al.*, Mass measurements of neutron-rich $A \approx 90$ nuclei constrain element abundances, Phys. Rev. C **109**, 035804 (2024).

[55] K.-L. Wang, A. Estrade, M. Famiano, H. Schatz, M. Barber, T. Baumann, D. Bazin, K. Bhatt, T. Chapman, J. Dopfer, B. Famiano, S. George, M. Giles, T. Ginter, J. Jenkins, S. Jin, L. Klankowski, S. Liddick, Z. Meisel, N. Nepal *et al.*, Mass measurements of neutron-rich nuclei near $N = 70$, Phys. Rev. C **109**, 035806 (2024).

[56] A. Jaries, S. Nikas, A. Kankainen, T. Eronen, O. Beliuskina, T. Dickel, M. Flayol, Z. Ge, M. Hukkanen, M. Mougeot, I. Pohjalainen, A. Raggio, M. Reponen, J. Ruotsalainen, M. Stryjczyk, and V. Virtanen, Probing the $N = 104$ midshell region for the $r$ process via precision mass spectrometry of neutron-rich rare-earth isotopes with the JYFLTRAP double penning trap, Phys. Rev. C **110**, 045809 (2024).

[57] S. Kimura, M. Wada, H. Haba, H. Ishiyama, S. Ishizawa, Y. Ito, T. Niwase, M. Rosenbusch, P. Schury, and A. Takamine, Comprehensive mass measurement study of $^{252}$Cf fission fragments with MRTOF-MS and detailed study of masses of neutron-rich ce isotopes, Phys. Rev. C **110**, 045810 (2024).

[58] C. M. Ireland, F. M. Maier, G. Bollen, S. E. Campbell, X. Chen, H. Erington, N. D. Gamage, M. J. Gutiérrez, C. Izzo, E. Leistenschneider, E. M. Lykiardopoulou, R. Orford, W. S. Porter, D. Puentes, M. Redshaw, R. Ringle, S. Rogers, S. Schwarz, L. Stackable, C. S. Sumithrarachchi *et al.*, High-precision mass measurement of $^{103}$Sn restores smoothness of the mass surface, Phys. Rev. C **111**, 014314 (2025).

[59] M. Mukai, Y. Hirayama, P. Schury, Y. X. Watanabe, T. Hashimoto, N. Hinohara, S. C. Jeong, H. Miyatake, J. Y. Moon, T. Niwase, M. Reponen, M. Rosenbusch, H. Ueno, and M. Wada, Evidence for shape transitions near $^{189}$W through direct mass measurements, Phys. Rev. C **111**, 014322 (2025).

[60] T. Wolfgruber, M. Knöll, and R. Roth, Precise neural network predictions of energies and radii from the no-core shell model, Phys. Rev. C **110**, 014327 (2024).

[61] E. Leistenschneider, E. Dunling, G. Bollen, B. A. Brown, J. Dilling, A. Hamaker, J. D. Holt, A. Jacobs, A. A. Kwiatkowski, T. Miyagi, W. S. Porter, D. Puentes, M. Redshaw, M. P. Reiter, R. Ringle, R. Sandler, C. S. Sumithrarachchi, A. A. Valverde, and I. T. Yandow (LEBIT Collaboration and TITAN Collaboration), Precision mass measurements of neutron-rich scandium isotopes refine the evolution of $N = 32$ and $N = 34$ shell closures, Phys. Rev. Lett. **126**, 042501 (2021).

[62] S. Beck, B. Kootte, I. Dedes, T. Dickel, A. A. Kwiatkowski, E. M. Lykiardopoulou, W. R. Plaß, M. P. Reiter, C. Andreoiu, J. Bergmann, T. Brunner, D. Curien, J. Dilling, J. Dudek, E. Dunling, J. Flowerdew, A. Gaamouci, L. Graham, G. Gwinner, A. Jacobs *et al.*, Mass measurements of neutron-deficient Yb isotopes and nuclear structure at the extreme proton-rich side of the $N = 82$ shell, Phys. Rev. Lett. **127**, 112501 (2021).

[63] H. F. Li, S. Naimi, T. M. Sprouse, M. R. Mumpower, Y. Abe, Y. Yamaguchi, D. Nagae, F. Suzaki, M. Wakasugi, H. Arakawa, W. B. Dou, D. Hamakawa, S. Hosoi, Y. Inada, D. Kajiki, T. Kobayashi, M. Sakaue, Y. Yokoda, T. Yamaguchi, R. Kagesawa *et al.*, First application of mass measurements with the Rare-RI ring reveals the solar $r$-process abundance trend at $A = 122$ and $A = 123$, Phys. Rev. Lett. **128**, 152701 (2022).

[64] Z. Ge, M. Reponen, T. Eronen, B. Hu, M. Kortelainen, A. Kankainen, I. Moore, D. Nesterenko, C. Yuan, O. Beliuskina, L. Cañete, R. de Groote, C. Delafosse, T. Dickel, A. de Roubin, S. Geldhof, W. Gins, J. D. Holt, M. Hukkanen, A. Jaries *et al.*, High-precision mass measurements of neutron deficient silver isotopes probe the robustness of the $N = 50$ shell closure, Phys. Rev. Lett. **133**, 132503 (2024).

[65] M. Zhang *et al.*, $B\rho$-defined isochronous mass spectrometry and mass measurements of $^{58}$Ni fragments, Eur. Phys. J. A **59**, (2023).

[66] M. Mougeot *et al.*, Mass measurements of $^{99–101}$In challenge *ab initio* nuclear theory of the nuclide $^{100}$sn, Nat. Phys. **17**, (2021).

[67] T. Otsuka, A. Gade, O. Sorlin, T. Suzuki, and Y. Utsuno, Evolution of shell structure in exotic nuclei, Rev. Mod. Phys. **92**, 015002 (2020).

[68] See https://www.researchgate.net/publication/389851541_FMTE_and_17_other_binding_energy_models_from_Machine_Learning for the binding energy values for FMTE and the other 16 models developed here.